\documentclass[12pt]{article}
\usepackage{oldgerm}
\usepackage{euler}
\usepackage{graphics}
\usepackage{bm}
\usepackage{graphicx}
\usepackage{amsmath}
\usepackage{amssymb}
\usepackage{amstext}
\usepackage{amscd}
\usepackage{amsfonts}
\usepackage{appendix}
\usepackage{color}

\DeclareMathAlphabet{\EuFrak}{U}{euf}{m}{n}
\DeclareMathAlphabet{\EuScript}{U}{eus}{m}{n}

\newcommand{\nd}{\noindent}

\newcommand{\be}{\begin{equation}}
\newcommand{\ee}{\end{equation}}
\newcommand{\ben}{\begin{eqnarray}}
\newcommand{\een}{\end{eqnarray}}

\title{{\bf Troublesome aspects of the Renyi-MaxEnt treatment}}

\author{{A. Plastino$^{3,5,6}$,
M.C.Rocca$^{3,4,5}$, F. Pennini$^{1,2}$}\\
\small{$^{1}$Universidad Cat\'olica del Norte, Av.~Angamos~0610, Antofagasta, Chile.}\\
\small{$^{2}$Facultad de Ciencias Exactas y Naturales,}\\
\small{Universidad Nacional de La Pampa, Peru 151, 6300 Santa Rosa,}\\
\small{La Pampa, Argentina}\\
\small{$^3$ Departamento de F\'{\i}sica,
Universidad Nacional de La Plata,}\\
\small{$^4$ Departamento de Matem\'{a}tica,
Universidad Nacional de La Plata,}\\
\small{$^5$ Consejo Nacional de Investigaciones Cient\'{\i}ficas
y Tecnol\'{o}gicas}\\
\small{(IFLP-CCT-CONICET)-C. C. 727, 1900 La Plata -
Argentina}\\\small{$^6$  SThAR - EPFL, Lausanne, Switzerland}}

\date{\today}

\begin{document}

\maketitle

\begin{abstract}

We study in great detail the possible existence of a
Renyi-associated thermodynamics, with negative results. In
particular, we uncover a hidden relation in the Renyi's
variational problem (MaxEnt). This relation connects the two
associated Lagrange multipliers (Canonical Ensemble) with the mean
energy $<U>$ and the Renyi parameter $\alpha$. As a consequence of
such relation,  we obtain anomalous Renyi-MaxEnt thermodynamic
results.

\nd PACS: 05.30.-d, 05.20-y, 05.70.-a


\end{abstract}

\newpage

\renewcommand{\theequation}{\arabic{section}.\arabic{equation}}

\setcounter{equation}{0}

\section{Introduction}

The Renyi information measure $S_R$ is a generalization of both
the Hartley and the Shannon ones, quantifying a system's
diversity, uncertainty, or randomness. $S_R$ is an important
quantity for several areas of scientific endeavor. One can
mention, for instance,  ecology, quantum information,  the
Heisenberg XY spin chain model, theoretical computer science,
conformal field theory, quantum quenching, diffusion processes,
etc. As a small sample, see for example,
\cite{1,2,3,4,5,6,7,8,9,10}.

\vskip 3mm \nd Information Theory (IT)  yields an extremely
powerful inference approach, usually abbreviated as MaxEnt
\cite{jaynes}. MaxEnt is able to describe quite general properties
of arbitrary systems, in several areas of Science,  on the basis
of scarce information. MaxEnt purports to provide one with  the
least-biased description that can be generated according to some
specific data, in any possible circumstances  \cite{jaynes}. In
the framework of statistical mechanics (SM), Jaynes  pioneered the
use of these IT ideas in order to both i) reformulate and ii)
generalize the SM-foundations  \cite{jaynes}. In this paper we
study Renyi-properties in a MaxEnt environment.

\vskip 3mm \nd  It is well known that Renyi's entropic functional
is not trace form. For all trace form functionals $F$, it has been
shown in \cite{univers} that they, together with  the MaxEnt
strictures, are able to reproduce the mathematical
Legendre-invariant structures of thermodynamics. Thus, one may
speak of an ``$F$-thermodynamics". Of course, this is not
guaranteed in the Renyi case, due to its lack of trace-class
nature. In this paper we carefully investigate further the issue
and conclude that there is no Renyi-associated thermodynamics. The
main culprit of this Renyi-failure is a hidden relation involving
the Renyi's MaEent-Lagrange multipliers that, as far as we know,
has not been discovered before.

\vskip 3mm \nd The paper is organized as follows: Section 2 deals
with the conventional Renyi's MaxEnt treatment and compares it
with Tsallis' one. Section 3 starts discovering some Renyi's
MaxEnt thermodynamic troubles, while Section 4 deals with the
hidden constraint referred to above. Section 5 illustrates our
ideas with reference to a simple, analytically tractable problem,
while some conclusions are drawn in Section 6.

\setcounter{equation}{0}

\section{Conventional MaxEnt Treatments}

\subsection{Renyi's MaxEnt}

Renyi's $S_R$ is defined as \cite{9}:
\begin{equation}
\label{eq2.1} S_R=\frac {1} {1-\alpha}\ln\left( \int\limits_M
P^{\alpha}d\mu\right),
\end{equation}
and the accompanying (canonical ensemble) MaxEnt probability
distribution $P$ arises from the maximization of the functional
$F_{S_R}(P)$ [where $U$ denotes the energy and $<U>$ its mean
value]
\begin{equation}
\label{eq2.2} F_{S_R}(P)=\frac {1} {1-\alpha}\ln\left(
\int\limits_M P^{\alpha}d\mu\right)+ \lambda_1\left(\int\limits_M
PUd\mu-<U>\right)+ \lambda_2\left(\int\limits_M Pd\mu-1\right),
\end{equation}
 Following
standard procedure  we consider the functional-$h$ increment
\cite{tp1,tp2}
\[F_{S_R}(P+h)=\frac {1} {1-\alpha}\ln\left[
\int\limits_M
(P+h)^{\alpha}d\mu\right]+
\lambda_1\left[\int\limits_M
(P+h)Ud\mu-<U>\right]+\]
\begin{equation}
\label{eq2.3} \lambda_2\left[\int\limits_M (P+h)d\mu-1\right],
\end{equation}
so that
\[F_{S_R}(P+h)-F_{S_R}(P)=\frac {1} {1-\alpha}\ln\left[
\int\limits_M
(P+h)^{\alpha}d\mu\right]-
\frac {1} {1-\alpha}\ln\left(
\int\limits_M
P^{\alpha}d\mu\right)+\]
\begin{equation}
\label{eq2.4} \lambda_1\int\limits_M hUd\mu+\lambda_2
\int\limits_M hd\mu.
\end{equation}
We now tackle $h^2$ contributions  so as to assess second
variations of $F_{S_R}$  \cite{tp1,tp2}
\[F_{S_R}(P+h)-F_{S_R}(P)=\frac {1} {1-\alpha}\ln\left\{
\int\limits_M
\left[P^{\alpha}+
\alpha hP^{\alpha-1}+
\frac {\alpha(\alpha-1)} {2}
h^2P^{\alpha-2}
d\mu\right]\right\}-\]
\begin{equation}
\label{eq2.5} \frac {1} {1-\alpha}\ln\left( \int\limits_M
P^{\alpha}d\mu\right)+ \lambda_1\int\limits_M hUd\mu+\lambda_2
\int\limits_M hd\mu,
\end{equation}
or,  equivalently,
\[F_{S_R}(P+h)-F_{S_R}(P)=\frac {1} {1-\alpha}\ln\left\{ 1+
\frac {\int\limits_M
\left[\alpha hP^{\alpha-1}+
\frac {\alpha(\alpha-1)} {2}
h^2P^{\alpha-2}
d\mu\right]}
{\int\limits_M
P^{\alpha} d\mu}
\right\}+\]
\begin{equation}
\label{eq2.6} \lambda_1\int\limits_M hUd\mu+\lambda_2
\int\limits_M hd\mu,
\end{equation}
so that one finally arrives at

\[F_{S_R}(P+h)-F_{S_R}(P)=\frac {1}
{1-\alpha} \frac {\int\limits_M \left[\alpha hP^{\alpha-1}+ \frac
{\alpha(\alpha-1)} {2} h^2P^{\alpha-2} d\mu\right]} {\int\limits_M
P^{\alpha} d\mu}+\]
\begin{equation}
\label{eq2.7} \lambda_1\int\limits_M hUd\mu+\lambda_2
\int\limits_M hd\mu.
\end{equation}
Summing up, we have for the first variation

\begin{equation}
\label{eq2.8} \frac {\alpha} {1-\alpha} \frac {P^{\alpha-1}}
{\int\limits_M P^{\alpha}d\mu}+\lambda_1 U+\lambda_2=0.
\end{equation}

\nd Functional calculus teaches that for the second variation one
must demand \cite{tp2}

\begin{equation} \label{eq2.9} -\alpha \frac
{\int\limits_M P^{\alpha-2}h^2 d\mu} {\int\limits_M
P^{\alpha}d\mu}\leq C||h||^2,
\end{equation}
with $C$ an arbitrary  negative constant \cite{tp2}.  {\it One
must remember that functional calculus is not identical to
ordinary calculus (involving ordinary functions), particularly
when one is looking for extremes \cite{tp1,tp2}.}\vskip 3mm

\nd The solution to  (\ref{eq2.8}) is ($Z$ below denotes Renyi's
partition function and $\beta$ the inverse temperature $1/T$)
\[\lambda_1=\beta(\alpha-1)\;\;\;;\;\;\;
\lambda_2=-1\;\;\;;\;\;\;\alpha<1\]
\begin{equation}
\label{eq2.10}
\lambda_1=\beta(1-\alpha)\;\;\;;\;\;\;
\lambda_2=1\;\;\;;\;\;\;\alpha>1
\end{equation}
\begin{equation}
\label{eq2.11}
Z=\int\limits_M
\left[1+(1-\alpha)\beta U\right]^{\frac {1}
{\alpha-1}}d\mu
\end{equation}
\begin{equation}
\label{eq2.12} P=\frac {1} {Z}
\left[1+(1-\alpha)\beta U\right]^{\frac {1} {\alpha-1}}
\end{equation}
As for the second variation we specialize things to quadratic
Hamiltonians (they are positive-definite) and restrict ourselves
to scenarios with  $\alpha<1$.
\[-\alpha
\frac {\int\limits_M
P^{\alpha-2}h^2 d\mu}
{\int\limits_M
P^{\alpha}d\mu}
\leq -\alpha
\int\limits_M
P^{\alpha-2}h^2 d\mu\leq\]
\[-\alpha Z^{2-\alpha}
\int\limits_M
\left[1+(1-\alpha)\beta U\right]^{\frac {\alpha-2}
{\alpha-1}}h^2d\mu\leq
-\alpha Z^{2-\alpha}
\int\limits_M
h^2 d\mu=\]
\begin{equation}
\label{eq2.13} -\alpha Z^{2-\alpha} ||h||^2\leq C||h||^2.
\end{equation}
It is clear that we can choose  $C$ in the fashion
\begin{equation}
\label{eq2.14} -\alpha Z^{2-\alpha}=C.
\end{equation}
\vskip 3mm  \nd Some new results emerge already at this level. We
see that for i) quadratic, positive definite Hamiltonians and ii)
$\alpha<1$, the MaxEnt functional $F_{S_R}$ attains always a
maximum. {\sf The novelty here resides in A) the restriction i)
and ii)} and B) for arbitrary Hamiltonians and $\alpha$'s,
(\ref{eq2.9}) must be investigated on a case-by-case basis.
Nothing can be stated a priori regarding the existence, or not, of
a MaxEnt maximum, contrary to popular belief.

\subsection{Comparison with Tsallis' MaxEnt}

During more than two decades, an important topic in statistical
mechanics theory revolved around the notion of generalized
q-statistics, pioneered by Tsallis \cite{t1}. It has been amply
demonstrated that, in many circumstances, the
Boltzmann-Gibbs-Shannon  logarithmic entropy does not yield a
correct description of the system under scrutiny \cite{t2}. Other
entropic forms, called q-entropies, produce a much better
performance \cite{t2}. One may cite a large number of such
instances. For example, non-ergodic systems exhibiting a complex
dynamics \cite{t2}. The non-extensive statistical mechanics of
Tsallis has been employed in many different areas of scientific
endeavor \cite{t3}.

\vskip 3mm  \nd Tsallis's entropic functional is both trace form
and a  monotonous function of $S_R$. The associated MaxEnt
functional reads \cite{t1}
\begin{equation}
\label{eq2.15} F_{S_T}(P)=-\int\limits_M\,
P^q\ln_q(P)\;d\mu+\lambda_1\left(
\int\limits_MPU\;d\mu-<U>\right)+\lambda_2
\left(\int\limits_MP\;d\mu-1\right),
\end{equation}
so that Tsallis' MaxEnt functional' first increment  becomes
\[F_{S_T}(P+h)-F_{S_T}(P)=-\int\limits_M(P+h)^q\ln_q(P+h)\;
d\mu+\lambda_1
\int\limits_MhU\;d\mu+\]
\begin{equation}
\label{eq2.16} \lambda_2\int\limits_Mh\;d\mu+
\int\limits_MP^q\ln_q(P)\;d\mu.
\end{equation}
The second order (in $h$)  for this MaxEnt functional is
\[F_{S_T}(P+h)-F_{S_T}(P)=
\int\limits_M\left[\left(\frac {q} {1-q}\right)P^{q-1}
+\lambda_1 U +\lambda_2\right]h\;d\mu-\]
\begin{equation}
\label{eq2.17} \int\limits_MqP^{q-2}\frac {h^2} {2}\;d\mu.
\end{equation}
From  (\ref{eq2.17}) we get
\begin{equation}
\label{eq2.18}
\left(\frac {q} {1-q}\right)P^{q-1} +\lambda_1 U
+\lambda_2=0,
\end{equation}
\begin{equation}
\label{eq2.19} -\int\limits_MqP^{q-2}h^2\;d\mu\leq C||h||^2.
\end{equation}
 The solution to (\ref{eq2.18}) is
\begin{equation}
\label{eq2.20} \lambda_1=-\beta q  Z^{1-q}_T,
\end{equation}
\begin{equation}
\label{eq2.21} \lambda_2=\frac {q} {q-1}Z^{1-q}_T,
\end{equation}
\begin{equation}
\label{eq2.22} P=\frac {[1+\beta(1-q)U]^{\frac {1} {q-1}}} {Z_T},
\end{equation}
\begin{equation}
\label{eq2.23} Z_T=\int\limits_M [1+\beta(1-q)U]^{\frac {1}
{q-1}}\;d\mu.
\end{equation}
Note that Eqs.  (\ref{eq2.9}) and (\ref{eq2.19}) differ just in a
constant. Consequently, Renyi's and  Tsallis' maxima coincide. For
a quadratic Hamiltonian we have
\[-\int\limits_MqP^{q-2}h^2\;d\mu=
-\int\limits_MqZ^{2-q}_T [1+\beta(1-q)U]^{\frac {q-2} {q-1}}
h^2\;d\mu\leq\]
\begin{equation}
\label{eq2.24} - qZ^{2-q}_T ||h^2||\leq C ||h^2||,
\end{equation}
\begin{equation}
\label{eq2.25} - qZ^{2-q}_T=C,
\end{equation}
so that, for  $q=\alpha$, the bound  $C$ is the same in the two
entropic instances.

\setcounter{equation}{0}

\section{Renyi's MaxEnt's thermodynamic troubles}

Let us express $S_R$ in terms of  $Z$ and $<U>$. To this end we
replace in Eq.  (\ref{eq2.8}) for the first variation:

\begin{equation}
\label{eq2.88} \frac {\alpha} {1-\alpha} \frac {P^{\alpha-1}}
{\int\limits_M P^{\alpha}d\mu}+\lambda_1 U+\lambda_2=0,
\end{equation}
the values of  $\lambda_1$ and $\lambda_2$ given by (\ref{eq2.10})
and for the $P$-expression (\ref{eq2.12}) [for $\alpha<1$].
\begin{equation}
\label{eq2.26} \frac {\alpha} {1-\alpha} \frac
{1+\beta(1-\alpha)U} {Z^{\alpha-1}\int\limits_M
P^{\alpha}d\mu}+\lambda_1 U+\lambda_2=0,
\end{equation}
\begin{equation}
\label{eq2.27} \frac {\alpha} {1-\alpha} \frac
{1+\beta(1-\alpha)U} {Z^{\alpha-1}\int\limits_M
P^{\alpha}d\mu}-\beta(1-\alpha) U-1=0.
\end{equation}
From the last relation one easily obtains
\begin{equation}
\label{eq2.28} \frac {\alpha Z^{1-\alpha}}
{1-\alpha}=\int\limits_M P^{\alpha}d\mu.
\end{equation}
Thus we have for $S_R$
\begin{equation}
\label{eq2.29} S_R=\ln Z+\frac {1} {1-\alpha}\ln\left( \frac
{\alpha} {1-\alpha}\right).
\end{equation}
Analogously, for $\alpha>1$ we find
\begin{equation}
\label{eq2.30} S_R=\ln Z+\frac {1} {1-\alpha}\ln\left( \frac
{\alpha} {\alpha-1}\right).
\end{equation}
We realize that in both instances i) $S_R$ does NOT explicitly
depend upon $<U>$ and ii) is not defined for  $\alpha\rightarrow
1$, both troublesome results. In particular, as we shall see in
great detail  below, one expects the (canonical ensemble) entropy
to be a sum of two terms. One of them contains de logarithm of the
partition function. The other is $\beta <U>$. This does  not
happen for $S_R$, according to Rq. (\ref{eq2.30}).

\vskip 4mm \nd Instead, for  Tsallis entropy we have from
(\ref{eq2.18}):
\begin{equation}
\label{eq2.31} \left(\frac {q} {1-q}\right)P^{q-1} -q\beta
Z^{1-q}_T U +\frac {q} {q-1}Z^{1-q}_T=0,
\end{equation}
that, multiplied by  $P$ yields

\begin{equation} \label{eq2.32} \frac {P^q} {1-q} -\beta Z^{1-q}_T
UP +\frac {P} {q-1}Z^{1-q}_T=0.
\end{equation}
The last ration can be recast as
\begin{equation}
\label{eq2.33} \frac {P^q-P} {1-q} -\beta Z^{1-q}_T UP +\frac {P}
{q-1}Z^{1-q}_T+\frac {P} {1-q}=0,
\end{equation}
that can be integrated to yield
\begin{equation}
\label{eq2.34} S_T -\beta Z^{1-q}_T<U> +\frac {1}
{q-1}Z^{1-q}_T+\frac {1} {1-q}=0,
\end{equation}
or, equivalently,
\begin{equation}
\label{eq2.35} S_T -\beta Z^{1-q}_T<U> -\frac {Z^{1-q}_T-1}
{1-q}=0,
\end{equation}
so that  $S_T$ becomes, invoking the so-called q-logarithm $\ln_q$
\cite{t2,t4},
\begin{equation}
\label{eq2.36} S_T=\ln_qZ_T+\beta Z^{1-q}_T<U>,
\end{equation}
which does exist in the limit  $q \rightarrow 1$, where we
encounter
\begin{equation}
\label{eq2.37} S=\ln Z_{BG}+\beta<U>,
\end{equation}
the usual thermodynamic  Boltzmann-Gibbs relation. This crucial
relationship that exists both in the BG and Tsallis cases cannot
be reproduced \`a la Renyi, which constitutes a new result.

\setcounter{equation}{0}

\section{The Hidden Renyi-MaxEnt Relation}

We have seen above that, in the MaxEnt framework,  both  Tsallis
and Renyi functionals display the same extremes. This is due to
the fact that  Renyi's functional monotonously depends on
Tsallis', as it is well known \cite{tp2}. However, these assertion
lose some strength if one studies more closely Eq. (\ref{eq2.8}),
that we repeat below:
\begin{equation}
\label{eq2.888} \frac {\alpha} {1-\alpha} \frac {P^{\alpha-1}}
{\int\limits_M P^{\alpha}d\mu}+\lambda_1 U+\lambda_2=0.
\end{equation}
 Indeed, multiplying  it  by  $P$ we find
\begin{equation}
\label{eq3.1} \frac {\alpha} {1-\alpha} \frac {P^{\alpha}}
{\int\limits_M P^{\alpha}d\mu}+\lambda_1 PU+\lambda_2P=0.
\end{equation}
Integrating now we are led to
\begin{equation}
\label{eq3.2} \frac {\alpha} {1-\alpha} +\lambda_1<U>+\lambda_2=0.
\end{equation}
This is an important result, showing that  $\lambda_1$ and
$\lambda_2$ are NOT  independent Lagrange multipliers, as MaxEnt
assumes. We are authorized  to write
\begin{equation}
\label{eq3.3} \lambda_2= \frac {\alpha} {\alpha-1} -\lambda_1<U>,
\end{equation}
and replacing this value of  $\lambda_2$ in (\ref{eq2.8}) we get

\begin{equation}
\label{eq3.4} \frac {\alpha} {1-\alpha} \frac {P^{\alpha-1}}
{\int\limits_M P^{\alpha}d\mu}+\lambda_1(U-<U>)+ \frac {\alpha}
{\alpha-1}=0,
\end{equation}
whose solution is given by

\begin{equation}
\label{eq3.5}
\lambda_1=-\beta\alpha
\end{equation}

\begin{equation}
\label{eq3.6}
P=\frac {[1+\beta(1-\alpha)(U-<U>)]^{\frac {1} {\alpha-1}}} {Z}
\end{equation}

\begin{equation}
\label{eq3.7}
Z=\int\limits_M [1+\beta(1-\alpha)(U-<U>)]^{\frac {1}
{\alpha-1}}\;d\mu.
\end{equation}
Using (\ref{eq3.6}), the second variation equation (\ref{eq2.9})
becomes
\begin{equation}
\label{eq3.8} -\int\limits_M\alpha P^{\alpha-2}h^2\;d\mu=
-\int\limits_M\alpha Z^{2-\alpha}
[1+\beta(1-\alpha)(U-<U>)]^{\frac {\alpha-2} {\alpha-1}}
h^2\;d\mu\leq C ||h^2||.
\end{equation}
At this stage, two important new results ensue. Contrarily to what
happened in Section 2, we cannot assert now that, for a quadratic,
positive-definite Hamiltonian, the Renyi functional exhibits a
MaxEnt maximum for $\alpha<1$. Even worse, within the MaxEnt
framework Renyi's expression is no longer a monotonous function of
the  Tsallis' one.

\nd Repeating now the steps of the preceding Section so as to
encounter a thermodynamic relation between $S_R$,  $Z$, and $<U>$
 we find, starting with (\ref{eq2.888}),
\begin{equation}
\label{eq3.9} \frac {P^{\alpha-1}} {\int\limits_M P^{\alpha}d\mu}+
\frac {1-\alpha} {\alpha} \lambda_1(U-<U>)-1=0,
\end{equation}

\begin{equation}
\label{eq3.10} \frac {P^{\alpha-1}} {\int\limits_M
P^{\alpha}d\mu}=1+ \frac {\alpha-1} {\alpha} \lambda_1(U-<U>)=0.
\end{equation}
Now we use (\ref{eq3.6}) to arrive at

\begin{equation}
\label{eq3.11} \frac { 1+ \frac {\alpha-1} {\alpha}
\lambda_1(U-<U>)} {Z^{\alpha-1}\int\limits_M P^{\alpha}d\mu}=1+
\frac {\alpha-1} {\alpha} \lambda_1(U-<U>)=0.
\end{equation}
From Eq. (\ref{eq3.11}) we get

\begin{equation} \label{eq3.12}
\int\limits_M P^{\alpha}d\mu=Z^{1-\alpha}
\end{equation}
\begin{equation}
\label{eq3.13} S_R= \frac {1} {1-\alpha}\ln\left( \int\limits_M
P^{\alpha}d\mu\right)= \frac {1} {1-\alpha} \ln{[Z^{1-\alpha}]},
\end{equation}
and, finally, the rather surprising relation

\begin{equation}
\label{eq3.14} S_R=\ln Z,
\end{equation}
an important new result. The essential link between statistical
mechanics and thermodynamics is the relation between the entropy,
$\beta <U>$, and the logarithm of the partition function, relation
that defines Helmholtz' free energy. This is lost here, entailing
that there is no Renyi-thermodynamics. \vskip 3mm

\nd Without the hidden constraint, the $S_R-$MaxEnt probabilities
 and  partition function are given by, respectively, Eqs.
 (\ref{eq2.11}) and (\ref{eq2.12}), which are the equations employed in the Literature. But the hidden constraint
 changes this situation to Eqs. (\ref{eq3.6}) and (\ref{eq3.7}),
 with devastating thermodynamic consequences.

\nd Further, from  (\ref{eq3.14}) we realize that $S_R$ does NOT
reduce to the Boltzmann-Gibbs entropy for $\alpha\rightarrow 1$.

\setcounter{equation}{0}

\section{Two-Levels Model for fixed $\alpha=q=2$}

As an illustration we consider a two-level model with $U_1=0$
, $U_2=1$, and $\alpha=q=2$. From (\ref{eq2.11}) and
(\ref{eq2.12}) we obtain:
\begin{equation}
\label{eq4.1}
Z= 2-\beta\;\;\;;\;\;\;
P_1=\frac {1} {2-\beta}\;\;\;;\;\;\; P_2=\frac {1-\beta}
{2-\beta}.
\end{equation}
and we get [see Eq. (\ref{eq2.11})]
\begin{equation}
\label{eq4.2} S_R=-\ln\left[ \left(\frac {1} {2-\beta}\right)^2+
\left(\frac {1-\beta} {2-\beta}\right)^2\right]
\end{equation}
\begin{equation}
\label{eq4.3} S_T=1-\left[ \left(\frac {1} {2-\beta}\right)^2+
\left(\frac {1-\beta} {2-\beta}\right)^2\right]
\end{equation}
From  (\ref{eq4.2}) and (\ref{eq4.3}) we see that $S_R$ and $S_T$
display the same maxima.

\vskip 3mm

 \nd Instead, if we consider Renyi's hidden
relation one must use Eqs. (\ref{eq3.6}) and (\ref{eq3.7}) to deduce the expressions
\begin{equation}
\label{eq4.4} Z=2+2\beta P_2-\beta,
\end{equation}
\begin{equation}
\label{eq4.5} P_1=\frac {1+\beta P_2} {2+2\beta P_2-\beta},
\end{equation}
\begin{equation}
\label{eq4.6} P_2=\frac {1+\beta P_2-\beta} {2+2\beta P_2-\beta},
\end{equation}
From (\ref{eq4.6}) we obtain a quadratic equation for  $P_2$, with
two solutions, one of which leads to a negative $P_2$ and becomes
inadmissible. Accordingly, for
\begin{equation}
\label{eq4.9} P_2^2+ \frac {(1-\beta)P_2} {\beta}+ \frac {\beta-1}
{2\beta}=0,
\end{equation}
 we are left with the solution
\begin{equation}
\label{eq4.10} P_2=\frac {\sqrt{\beta^2-1}+\beta-1} {2\beta},
\end{equation}
so that, after suitable replacement, we obtain
\begin{equation}
\label{eq4.11} P_1=\frac {\sqrt{\beta^2-1}+\beta+1}
{2\sqrt{1-\beta^2}+1}.
\end{equation}
Finally, the entropy becomes
\begin{equation}
\label{eq4.12} S_R=-\ln\left[\left(\frac
{\sqrt{\beta^2-1}+\beta+1} {2\sqrt{1-\beta^2}+1}\right)^2+
\left(\frac {\sqrt{\beta^2-1}+\beta-1} {2\beta}\right)^2\right],
\end{equation}
which i) it is not a monotone function of Tsallis' entropy, and
ii) it does not display the  Tsallis' maxima.

\setcounter{equation}{0}

\section{Conclusions}


We studied in great detail the possible existence of a Renyi's
thermodynamics, with negative results. Summing up:

\begin{itemize}

\item As a first result we saw that for i) quadratic, positive
definite Hamiltonians and  ii) $\alpha<1$, the MaxEnt functional
$F_{S_R}$ attains always a maximum. {\sf The novelty here resides
in}

\begin{enumerate}

\item point i) above and the  $\alpha-$restriction  ii)

\item  for arbitrary Hamiltonians and $\alpha$'s,
(\ref{eq2.9}) must be investigated on a case-by-case basis.
Nothing can be stated a priori regarding the existence, or not, of
a MaxEnt maximum, contrary to popular belief.
\end{enumerate}

\item $S_R$ does NOT explicitly depend upon $<U>$ and is not defined
for $\alpha\rightarrow 1$, both troublesome results.

\item The relation
\begin{equation}
\label{eq2.371} S=\ln Z_{BG}+\beta<U>,
\end{equation}
is a crucial thermodynamic  Boltzmann-Gibbs relation. This
critical relationship that exists both in the BG and Tsallis cases
cannot be reproduced \`a la Renyi [because $S_R=\ln{Z}$], which
constitutes a new result.

\item The hidden $S_R-$MaxEnt relation
\begin{equation}
\label{eq3.31} \lambda_2= \frac {\alpha} {\alpha-1} -\lambda_1<U>,
\end{equation}
linking $\alpha$, $<U>$,  and the two Lagrange multipliers, is a
crucial new result.

\item As a consequence, contrarily to
what happened in Section 2, we cannot assert  that, for a
quadratic, positive-definite Hamiltonian, the Renyi functional
exhibits a MaxEnt maximum for $\alpha<1$. Even worse, within the
MaxEnt framework Renyi's expression is no longer a monotonous
function of the  Tsallis' one. Without the hidden constraint, the
$S_R$'s MaxEnt probabilities
 and  partition function are given by, respectively, Eqs.
 (\ref{eq2.11}) and (\ref{eq2.12}), which are the equations employed in the Literature. But the hidden constraint
 changes this situation to Eqs. (\ref{eq3.6}) and (\ref{eq3.7}),
 with devastating thermodynamic consequences.
\end{itemize}

Finally, let us insist: the essential link between statistical
mechanics and thermodynamics is the relation between the entropy,
$\beta <U>$, and the logarithm of the partition function, relation
that defines Helmholtz' free energy. This is lost here, entailing
that there is no Renyi-thermodynamics.


\begin{thebibliography}{99}
\bibitem{t4} The q-logarithm is defined in the fashion \cite{t2}  $\ln_q Z= \frac {Z^{1-q}-1}
{1-q}$.
\end{thebibliography}
\end{document}